\documentclass{article}

\usepackage{arxiv}

\usepackage[utf8]{inputenc}
\usepackage[french, english]{babel}
\usepackage[T1]{fontenc}
\usepackage{amsmath}
\usepackage{amsfonts}
\usepackage{amssymb}
\usepackage{graphicx}
\usepackage[figurename=Fig.]{caption}
\usepackage{subcaption}
\usepackage{thmtools, thm-restate}
\usepackage{hyperref}
\usepackage{comment}
\usepackage{amsthm}
\usepackage[titletoc]{appendix}
\usepackage{xcolor}
\usepackage{caption}

\newtheorem{theorem}{Theorem}[section]

\newtheorem{lemma}[theorem]{Lemma}

\newtheorem{defn}[theorem]{Definition}
\newtheorem{rem}[theorem]{Remark}
\title{Mathematical analysis of a delayed  SEIRDS epidemics models: deterministic and stochastic approach}

\author{
 Mohamed Ben Alaya \\
  Laboratoire de Mathématiques\\
  Raphaël Salem UMR 6085 CNRS-Université\\
  de Rouen Normandie, France
   \And
 Walid Ben Aribi\\
  University of Tunis El Manar\\
  Pasteur Institute of Tunis\\
  LR16IPT09   Bio-(Informatics, Mathematics and Statistics)\\
  Lab, Tunis, Tunisia\\
  \texttt{ben.aribi.walid89@gmail.com} \\
  \And
 Slimane Ben Miled \\
  University of Tunis El Manar\\
  Pasteur Institute of Tunis\\
  LR16IPT09   Bio-(Informatics, Mathematics and Statistics)\\
  Lab , Tunis, Tunisia
}

\begin{document}
\maketitle
\begin{abstract}
The primary goal of this research is to investigate the impact of delay on the dynamics of the Susceptible-Exposed-Infected-Recovered-Death and Susceptible (SEIRDS) model, to which we add a stochastic term to account for uncertainty in COVID-19 parameter estimations.
We run two models, one deterministic and one stochastic, and show that their solutions exist and are unique.
We also numerically investigate the impact of immunity loss on the emerging time of a new wave, as well as the necessary condition for the extinction and persistence of the disease.
\end{abstract}

\keywords{Stochastic epidemic model \and SEIRDS \and Delay \and loss of immunity}

\section{Introduction}	
A new coronavirus epidemic has erupted throughout the world since the end of December 2019 killing almost 6 million people worldwide\footnote{https://www.worldometers.info/coronavirus/}.

Soon, hopes turned to the need to vaccinate as many people as possible \cite{ref48}. 
We rapidly saw a decrease in the number of positive cases at the start of the vaccination program \cite{ref47}.
However, one year later we observed  a rebound in the epidemic, even in countries with extremely high vaccination rates  \cite{ref34,ref38}.
This resurgence is triggered by a lack of immunity in COVID-19 individuals or by cross-immunity, which makes people vulnerable to new variants \cite{ref4,ref5,ref12,ref16,ref37}.

The first epidemiological model, the  SIR (Susceptible-Infectious-Recovery) Model, was presented by Kermack-McKendrick in 1927. This model made the assumption that the population is divided into three compartments,  Susceptible, Infected, and Recovered and  lifetime immunity to the disease  \cite{ref45,ref46}.
Many important extensions were then been developed from the classical SIR model to a more complicated model making epidemic modeling more realistic. 

COVID-19 is a disease in which individuals with no symptoms can carry the virus, especially in the early stages of infection.
This feature necessitates the extension of the classical SIR model by the inclusion of a new  compartment: the exposed, $E$. We thus obtain,  SEIR/DS (Susceptible-Exposed-Infectious-Recovery-Dead) \cite{ref7,ref36}.
Moreover, the loss of immunity lead to a delay in the transition from the recovery compartment to the susceptible one.


Epidemic models with delays, were recently used by  Hethcote and van der Driessche for modeling  infectious duration \cite{ref14a} or immunity loos \cite{ref14b}. Note that  Cooke and Van Den Driessche \cite{ref3a} used time delays to present the latency and temporary immunity periods in an SEIRS model. 


COVID-19 outbreak exhibits significant regional and temporal variability  \cite{ref43,ref44}  caused by complex social relations and interactions with public health decisions. These variabilities can be taken into account by random perturbation of model parameters \cite{ref31,ref32,ref24,ref8}.
Some models used Markov time method chains \cite{ref10}, or  Lévy jump noise \cite{ref4a,ref42}, 
Similar ideas have also been used in  SEIR model without delay \cite{ref14,ref6,ref18,ref27}, and with delay \cite{ref2a,ref4a,ref4b,ref26,ref35,ref23}.




In this work, we propose a deterministic epidemic model SEIR/DS (Susceptible-Exposed-Infected-Recovered-Death-Susceptible) for COVID-19 with a delay representing the loss of immunity. We extend this model by considering a stochastic differential equation with delay while adding the noise term in the rate of transmission and by adding stochastic perturbations proportional to $S$, $E$, and $I$. We prove  results, on the existence and uniqueness of the solution and on the asymptotic behavior of the solution.
All results are for deterministic and for stochastic delay equations.

The document is organized as follows: In section \ref{sec:2}, the formulation of a deterministic SEIRDS model is presented. We discuss the case with and without delay. In section \ref{sec:3} we study the stochastic model. Numerical simulations are given in section \ref{sec:4}. Finally, the conclusion is given in section \ref{sec:5}.

\section{Deterministic modeling of SEIRDS Model}
\label{sec:2}
To study the spread of COVID-19 disease, we consider the SEIR/DS model taking into account the loss of immunity. Where, $S$, the "Susceptible" state characterizes not infected individuals who live in an environment where the virus circulates, $E$, the “Exposed” state characterizes contaminated and infectious individuals, who are in the early stages of infection but not yet symptomatic (incubation period), 
$I$ the "infected" state characterizes infected and infectious  individuals, which can be  symptomatic  or asymptomatic, $R$ the "recovered" state characterizes the individuals who are no longer infected and are immunized 
$D$ the "Deceased" state represents the individuals who died as a result of the disease.

\begin{figure}
    \centering
    \includegraphics{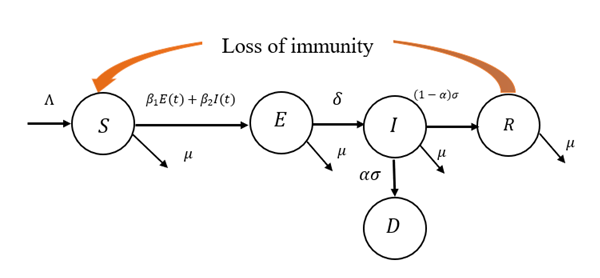}
    \caption{The flow diagram of the COVID-19 infection.}
    \label{fig:00}
\end{figure}

We suppose that a susceptible individual, $S$, becomes exposed, $E$,  after positive contact with an infected  individual at the early stage of infection, $E$,  at rate $ \beta_1 $ or with an infected individual, $I$, at a rate $\beta_2$. An exposed individual develops symptoms after an incubation period $ \frac{1}{\delta}$. An infected individual  die after $ \frac{1}{\sigma }$ days with the probability $\alpha$ or, cured with probability $(1-\alpha)$. Cured individuals lost their immunity and become susceptible at rate $\varphi$. Lets
$\mu $ is the natural death rate and $\Lambda$ the newborns (see figure \ref{fig:00}). All parameter values are assumed to be non-negative.

We will also assume that the incidence rate depends on the number of susceptible and infectious individuals at a given time $t$ (ie individuals in the incubation period can transmit the disease). 

Let denote for each time $t$, $S(t)$, $E(t)$, $I(t)$ and $R(t)$ the density of susceptible, exposed, infected, recovered (with temporary immunity acquired from a disease) individuals. We add a death class $D(t)$  to represent the individuals who died as a result of the disease. 
 The total population density is denoted by, $ N(t)=S(t)+E(t)+I(t)+R(t)$.


Then, the epidemiological model can be written as follows: 
\begin{equation}
\left\lbrace
\begin{aligned}
\frac{dS(t)}{dt} &= \Lambda-(\beta_1E(t)+\beta_2I(t))S(t) + \varphi R(t)- \mu S(t)\\
\frac{dE(t)}{dt} &= \beta_1S(t)E(t)+\beta_2S(t)I(t) - \left(\delta+\mu \right) E(t)\\
\frac{dI(t)}{dt} &= \delta E(t)- \left(\sigma+\mu \right) I(t)\\
\frac{dR(t)}{dt} &= \left(1-\alpha\right)\sigma I(t) - (\varphi +\mu )R(t)\\
 \frac{dD(t)}{dt} &= \alpha\sigma I(t).
\end{aligned}
\right.
\label{eq:0}
\end{equation}
with initial condition $ (S(0),~ E(0),~ I(0),~ R(0),~ D(0))\in \mathbb{R}_+^5 .$

The following theorem ensures that  the equation's \eqref{eq:0} solution exists and is unique.

\begin{theorem} \label{TH}
The system (\ref{eq:0}) admits a unique solution and the solution remains  in,
\begin{equation*}
\Gamma = \{\left(S,E,I,R,D \right) \in \mathbb{R}^5 | S\geq 0,E\geq 0,I\geq 0,R\geq 0,D\geq 0, ~S+E+I+R \leq \frac{\Lambda}{\mu} \}.   
\end{equation*}
\end{theorem}

\begin{proof} 
First, the variable $D$ does not appear in the four first  equations so it is sufficient to analyze the behavior of the solutions of the four first equations of the model (\ref{eq:0}). 

Let $X(t) = (S(t),E(t),I(t),R(t))$ then the system \eqref{eq:0} without the last equation can be writing as $\dot{X}(t)=F(X(t))$, where $F \in C^{1}(\mathbb{R}_{+}^4,\mathbb{R})$. Since $F$ is locally Lipschitzian then there exists a unique maximal solution to the problem of Cauchy Lipschitz associated with our differential system for an initial condition $(S(0),~ E(0),~ I(0),~ R(0)) \in \mathbb{R}_{+}^4 .$ 

Furthermore, we have,
$\frac{dS(t)}{dt}|_{S(t)=0}=\Lambda +\varphi R(t)\geq 0$, $\frac{dE(t)}{dt}|_{E(t)=0}=\beta_2 S(t)I(t) \geq 0$, $\frac{dI(t)}{dt}|_{I(t)=0}=\delta E(t) \geq 0$ and $\frac{dR(t)}{dt}|_{R(t)=0}= \left(1-\alpha\right)\sigma I(t) \geq 0$. Then the solution of (\ref{eq:0}) will be a positive (see, for example \cite[proposition 1]{ref22}). In addition, thanks to the definition of $\Gamma$, we have
$$
\Lambda - \mu N(t) - \alpha \sigma N(t)\leq \frac{dN(t)}{dt} = \Lambda - \mu N(t) - \alpha \sigma I(t) \leq \Lambda - \mu N(t).
$$
Then, according to the comparison theorem (\ref{LEM2}), we obtain,
\begin{equation}
\frac{\Lambda}{\alpha \sigma +\mu}+\left(N(0)-\frac{\Lambda}{\alpha \sigma +\mu}\right) e^{-(\alpha \sigma +\mu) t} \leq N(t) \leq \frac{\Lambda}{\mu}+\left(N(0)-\frac{\Lambda}{\mu}\right) e^{-\mu t}
\end{equation}

If $\frac{\Lambda}{ \alpha \sigma +\mu}\leq N(0) \leq \frac{\Lambda}{\mu}$, then,

\begin{equation}
\label{S}
\begin{aligned}
0 \leq \frac{\Lambda}{ \alpha \sigma +\mu}\leq N(t) \leq \frac{\Lambda}{\mu}.
\end{aligned}
\end{equation}

Then, the solution of the system (\ref{eq:0}) without the last equation remains in the positive region $\mathbb{R}_+^4$. Hence, the solution of the system (\ref{eq:0}) remains in the positive region $\Gamma$ defined as follows,

\begin{equation*}
\Gamma = \{\left(S,E,I,R,D \right) \in \mathbb{R}^5 | S\geq 0,E\geq 0,I\geq 0,R\geq 0,D\geq 0, ~S+E+I+R \leq \frac{\Lambda}{\mu} \} .   
\end{equation*}

Finally, by the boundedness of $S(t),E(t),I(t)$ and $R(t)$ we deduce that we have a global solution.  
\end{proof}

Following \cite{ref39}, the corresponding basic reproduction number, $R_0 $, of system (\ref{eq:0}),  takes the form,
\begin{equation}
\label{R0}
R_0 = \frac{\Lambda (\beta_1 (\sigma+\mu)+\beta_2 \delta)}{\mu \left(\delta+\mu\right)\left(\sigma  + \mu\right)}=\frac{\Lambda \beta_1}{\mu \left(\delta+\mu\right)}+\frac{\Lambda \beta_2 \delta}{\mu \left(\delta+\mu\right)\left(\sigma  + \mu\right)}, 
\end{equation}

which represent the average number of secondary transmissions from a single infectious individual in a fully susceptible population and correspond to the dominant eigenvalue of the next-generation matrix $ -F V^{-1}$, where $F=
\begin{pmatrix}
\beta_1 \frac{\Lambda}{\mu} & \beta_2 \frac{\Lambda}{\mu}\\
0 & 0 
\end{pmatrix}$ and $V=
\begin{pmatrix}
-\left(\delta+\mu\right) & 0 \\
\delta & -\left(\sigma+\mu\right) 
\end{pmatrix}.$

\begin{rem} 
\label{REM}
The system (\ref{eq:0}) admits a disease-free equilibrium $\left(\frac{\Lambda}{\mu},0,0,0\right) $ which exists for all parameter values. The endemic equilibrium $\left( \frac{\Lambda}{\mu R_0},~ \frac{\left(\sigma+\mu\right)}{\delta}I^*,~ I^*,~ \frac{\left(1-\alpha\right)\sigma}{\varphi+\mu}I^*\right)$ with $I^* = \frac{\Lambda \mu (R_0-1)\left(\varphi+\mu\right)\delta}{\Lambda \left(\beta_1(\sigma+\mu)+\beta_2\delta\right)(\varphi+\mu)-\varphi\left(1-\alpha\right)\mu  \sigma\delta R_0},
$
exists if $1 < R_0 < \frac{\Lambda \left(\beta_1(\sigma+\mu)+\beta_2\delta\right)(\varphi+\mu)}{\varphi\left(1-\alpha\right)\mu \sigma\delta }.$

\end{rem}

For many respiratory infections, like COVID-19, immunity to reinfection is not lifelong, and individuals can lose their immunity and become susceptible again \cite{ref5,ref37}. Therefore, after a period of immunity noted, $ \tau $ (in days), the recovered individual returns to the susceptible class, with the rate $ \varphi $. Indeed, $\tau$ is defined by the period between the time $t - \tau$ when an individual becomes immune after a period of contagiousness and the time $t$ when he starts to lose his immunity. At the time $\tau$, after recovery, hosts that did not die in the time interval $\tau$ become susceptible again. In other words, if an individual that recovers at time $t_1$, $R(t_1)>0$, survives to time $t_1 + \tau$, it leaves class $R$ and enters class $S$. In turn, we find a lag term in the equation for S as well, and we have a model with a constant lag. Moreover, if the loss of immunity occurs at time $t$ for hosts that have recovered from infection at time $t-\tau$, $R(t-\tau)$, and in the time interval $[t - \tau, t]$ the immune individual has survived (time $t-\tau$ presents the time that an individual enters the recovered compartment), then,  afterward $ R(t-\tau)e^{-\mu\tau} $ represents the number of individual who loses immunity and become susceptible again (The term $e^{-\mu\tau}$ describes the survival rate of  the immunized population in the period $\tau$). Then, the epidemic model with discrete delay is given by:

\begin{equation}
\left\lbrace
\begin{aligned}
\dot{S}(t) &= \Lambda -\beta_1S(t)E(t)-\beta_2S(t)I(t) + \varphi  R(t-\tau)e^{-\mu\tau}- \mu S(t)\\
\dot{E}(t) &= \beta_1S(t)E(t)+\beta_2S(t)I(t) - \left(\delta+\mu \right) E(t)\\
\dot{I}(t) &= \delta E(t)- \left(\sigma+\mu \right) I(t)\\
\dot{R}(t) &= \left(1-\alpha\right)\sigma I(t) -  \varphi R(t-\tau)e^{-\mu\tau} -\mu R(t)\\
\dot{D}(t) &= \alpha\sigma I(t)
\end{aligned}
\right.
\label{eq:1}
\end{equation}

with initial conditions $S(\theta)=\Phi_1(\theta)> 0,~ E(\theta)=\Phi_2(\theta)> 0,~I(\theta)= \Phi_3(\theta)> 0,~ R(\theta)=\Phi_4(\theta) > 0,~ D(\theta)= \Phi_5(\theta)> 0$, $\forall  \theta \in \left[-\tau, 0 \right]$, where, the function $\Phi_1,\Phi_2,\Phi_3,\Phi_4 \text{~and~} \Phi_5  \in C\left([-\tau, 0], \mathbb{R}^{5}_+\right)$, where $C\left([-\tau, 0], \mathbb{R}^{5}_+\right)  $ is the set of non negative real continuous function from $\left[-\tau, 0\right]$ into $\mathbb{R}_+^5$.
`
\begin{lemma} 
\label{lemma0}
Assume that $0 < \varphi \tau < e^{-1}$ then there exists a unique $\lambda$ satisfying the characteristic equation $\lambda = - \mu - \varphi e^{-\tau (\lambda+\mu)}$ and the following ordinary differential equation with delay

\begin{equation}
\label{eq:U}
 \left\lbrace
\begin{aligned}  
\dot{U}(t)&=-\mu U(t) - \varphi  U(t-\tau)e^{-\tau \mu},~\forall~t\geq 0\\
U(t) &=  C e^{\lambda t},~\forall~t\in [-\tau,0],~C \in \mathbb{R}
\end{aligned}
\right.
\end{equation}
has a unique nontrivial solution $Ce^{\lambda t}$.
\end{lemma} 
\begin{proof}
First, let $f(\lambda)=\lambda + \mu + \varphi e^{-\tau (\lambda+\mu)}$, under the condition $0 < \varphi  < \frac{e^{-1}}{\tau}$ we get 
$ f(-\mu - \frac{1}{\tau})=-\frac{1}{\tau}+ \varphi e < 0$, $ f(-\mu + \frac{1}{\tau})=\frac{1}{\tau}+ \varphi e >0$ and for all $ \lambda \geq -\mu -\frac{1}{\tau} $, $f'(\lambda)=1  -\varphi \tau  e^{-\tau (\lambda+\mu)}> 1  -\varphi \tau  e >  0$. Then there exists a unique $\lambda \in \left(-\mu - \frac{1}{\tau},-\mu + \frac{1}{\tau}\right) $ such that $f(\lambda) =0$.
It is straightforward that $C e^{\lambda t}$ is the a solution of equation \ref{eq:U}, with  $\lambda \in \left(-\mu - \frac{1}{\tau},-\mu + \frac{1}{\tau}\right) $. The uniqueness is given by Cauchy Lipschitz argument.
\end{proof}

\begin{theorem}
\label{thm : 2.4}
Under the condition $0 < \varphi \tau < e^{-1}$, 
the solution $(S(t), E(t),I(t), R(t),D(t))$ of (\ref{eq:1}) exists and is unique and positive for $t \geq -\tau$.
\end{theorem}
\begin{proof}
The variable D does not appear in the first four equations of the system, then, to study the existence and uniqueness of the solution it is sufficient to consider only the first four equations of the model. Let's $X(t)=\left(S(t), E(t), I(t), R(t)\right) \in \mathbb{R}_{+}^4$.
Designed the norm of  $\Phi \in C\left([-\tau, 0], \mathbb{R}^{4}_+\right)$ by   $\|\Phi\|=\max _{\theta \in[-\tau, 0]}|\Phi(\theta)|$, hence $ C\left([-\tau, 0], \mathbb{R}^{4}_+\right)$ becomes a Banach space. 
If $X(t)$ is continuous in $\left[-\tau, T\right)$ with $T>0$, for $0 \leqslant t<T$, we define $x_{t}=X(t+\theta),~ \theta \in[-\tau, 0]$. Then a system of delay-differential equations can be written as
\begin{equation}\label{xt}
    \left\lbrace
    \begin{aligned}
    &\dot{X}(t)=f\left(x_{t}\right)  \\
    &x_{0}=\Phi
    \end{aligned},
    \right.    
\end{equation}
with $$f\left(\Phi\right) = F(\Phi(0),\Phi(-\tau)) = 
\begin{pmatrix}
\Lambda -\beta_1 \Phi_1(0)\Phi_2(0)-\beta_2 \Phi_1(0)\Phi_3(0) + \varphi  \Phi_4(-\tau)e^{-\mu\tau}- \mu \Phi_1(0)\\
\beta_1 \Phi_1(0)\Phi_2(0)+\beta_2 \Phi_1(0)\Phi_3(0) - \left(\delta+\mu \right) \Phi_2(0)\\
\delta \Phi_2(0)- \left(\sigma+\mu \right) \Phi_3(0)\\
\left(1-\alpha\right)\sigma \Phi_3(0) -  \varphi \Phi_4(-\tau)e^{-\mu\tau} -\mu \Phi_4(0)
\end{pmatrix}.$$
It is clear that $f$ is locally Lipschitzian. It follows the theorems (\ref{THE3}), the solution $X(t)$ of (\ref{eq:1}) exists and is unique on $[-\tau , T)$ for some $T>0$. Now, as $\forall t \in [-\tau,0] $ $\Phi_i \geq 0,~i=1,2,3$ and $\Phi_4>0$, let us show that the solution $X(t)$ of (\ref{eq:1}) is positive for all $t\in [0, T)$ for some $T>0$. In fact, if $X(t)$ were to lose its non-negativity on $[0, T)$ then there would exist an instant $t_1 \in [0,T)$ such that $S(t_1)E(t_1)I(t_1)R(t_1)=0$ and $S(t)>0, E(t)>0, I(t)>0, R(t)>0,~\forall~t\in [0, t_1)$. Now we proceed by the absurd. If we assume that $S(t_{1}) = 0$ then by the first equation of the system  (\ref{eq:1}) we have $\dot{S}\left(t_{1}\right)=\Lambda+\varphi R(t_1-\tau)e^{-\mu\tau}>0$  ($t_1-\tau<t_1$)  and so $S(t)<0$, for all $ t \in] t_{1}-\varepsilon, t_{1}[$, where $\varepsilon>0$ is sufficiently small, which is in contradiction, and yields $S(t)\geq 0~ \forall t \in\left[0, T)\right.$. Now, integrating the third equation of (\ref{eq:1}) from 0 to $t_{1}$, we see that $
I\left(t_{1}\right)=I(0) e^{-\left(\sigma+\mu\right) t_{1}}+\int_0^{t_1} \delta E(r) e^{-(\sigma+\mu)(t_1-r)}>0
$, which is in contradiction with $I\left(t_{1}\right)=0$, then $I(t)\geq 0,~\forall~t~\in\left[0, T)\right.$. Using these results in the second equation of the system (\ref{eq:1}) and $E\left(t_{1}\right) = 0$, we get  $\dot{E}(t_1) = \beta_2 S(t_1)I(t_1) >0$ similarly as above we get $E(t)\geq 0,~\forall~t~\in\left[0, T)\right.$.

Now, $\forall~t\in [0, T)$ we have $I(t) \geq 0$, then $\dot{R}(t) \geq -\varphi R(t-\tau)e^{-\mu \tau} - \mu R(t)$, as $\Phi_4 > 0 $, then there exist $C > 0$ such that $\Phi_4(t) \geq Ce^{\lambda t},~\forall t\in [-\tau,0]$, where $\lambda$ is the solution of the characteristic equation $\lambda = - \mu - \varphi e^{-\tau (\lambda+\mu)}$ of the ordinary differential equation with delay 
$\dot{U}(t)= -\mu U(t) -\varphi U (t-\tau) e^{-\mu \tau}$. 
According to the lemma \eqref{lemma0} and comparison theorem, we get for all $t \in [0, T) $, $R(t) \geq Ce^{\lambda t} > 0$, which is in contradiction with $R (t_1) = 0$, then $R(t) \geq 0, \forall t \ [0, T )$. Consequently, $ \forall t \in[0, T)$, the solution of (\ref{eq:1}) is positive. Thus, using (\ref{S}), for $t \in[0, T)$, $$
\frac{\Lambda}{ \alpha \sigma +\mu}\leq N(t) \leq \frac{\Lambda}{\mu}
$$
which implies that $(S(t), E(t), I(t), R(t))$ is bounded on $[0, T)$.Then, it follows from (\cite{ref36a}, theorem 3.2 p26) that we have a global solution of \eqref{eq:1}. Finally, the there exists a unique and positive solution of \eqref{eq:1} on $[0,+\infty[$. This completes the proof of the theorem.
\end{proof}

%

\section{The dynamic behaviors of the stochastic model}
\label{sec:3}
In this section, we assume that the transmission coefficient $\beta_1$ depends on individual or group behavior \cite{ref11}, represented in our model  by random  perturbation on $\beta_1$ and $\beta_2$.  Let,
$ \left(\Omega,\mathcal{F},(\mathcal{F}_t)_{t\geq0},P\right) $ be a complete probability space with a filtration $(\mathcal{F}_t)_{t\geq0}$ satisfying the usual conditions, we replace  $\beta_1 dt ~\text{by}~ \beta_1dt+\eta_1 dW_1(t)$. Therefore the stochastic extend model \eqref{eq:1} is defined by,

\begin{equation}\label{eq:3}
\left\lbrace
\begin{aligned} 
dS(t)&=\left(\Lambda+ \varphi R(t-\tau)e^{-\mu\tau}- \beta_1S(t)E(t)-\beta_2S(t)I(t) -\mu S(t)\right) dt-\eta_1 S(t)E(t)dW_1(t) + \eta_2 S(t)dW_2(t) \\
dE(t) &=\left(\beta_1S(t)E(t)+\beta_2S(t)I(t)-\left(\delta +\mu \right) E(t)\right)dt + \eta_1 S(t)E(t) dW_1(t)+ \eta_3 E(t)dW_3(t)\\
d{I}(t) &=\left(  \delta E(t)- \left(\sigma+\mu\right) I(t)\right) dt+ \eta_4 I(t)dW_4(t) \\
d{R}(t) &=\left( \left( 1-\alpha\right)\sigma I(t)- \varphi  R(t-\tau)e^{-\mu\tau}-\mu R(t)\right) dt \\
d{D}(t) &= \alpha \sigma I(t) dt
\end{aligned}
\right.
\end{equation}

where $W_1(t)$, $W_2(t)$, $W_3(t)$ and $W_4(t)$ are standard Brownian motions, $\eta_1 > 0$, $\eta_2 > 0$, $\eta_3 > 0$ and $\eta_4 > 0$  are the intensities of white noises, and the initial conditions $S(\theta)=\Phi_1(\theta)> 0,~ E(\theta)=\Phi_2(\theta)> 0, I(\theta)= \Phi_3(\theta)> 0,~ R(\theta)=\Phi_4(\theta) > 0,~D(\theta)= \Phi_5(\theta)> 0$, with the functions $\Phi_1,\Phi_2,\Phi_3,\Phi_4 \text{~and~} \Phi_5  \in C\left(\left[-\tau, 0\right] ,\mathbb{R}_+^5\right)$. Then, the equation (\ref{eq:3}) can be written as follows:

\begin{equation}\label{eq:4}
\begin{aligned}
dX(t)&=F\left(t,X(t),X(t-\tau))dt+G(t,X(t),X(t-\tau)\right)dW(t),\\
\end{aligned}
\end{equation}

where $X(t)=\left( S(t), E(t), I(t), R(t), D(t)\right)\in\mathbb{R}^5$, $W(t)=(W_1(t),W_2(t),W_3(t),W_4(t)) \in \mathbb{R}^4 $,  the functions $F : \mathbb{R}_+  \times \mathbb{R}^{5} \times \mathbb{R}^{5}  \to\mathbb{R}^5$ and $G :\mathbb{R}_+  \times \mathbb{R}^{5} \times \mathbb{R}^{5} \to \mathbb{R}^{5}\times \mathbb{R}^{4}  $.
If for $T>0$ we write our system as a solution of the stochastic functional differential equation

\begin{equation}\label{eq:41}
\left\lbrace
\begin{aligned}
d x(t)&=f\left(t,x_{t}\right) d t+g\left(t,x_{t}\right) d W(t) \quad \text { on } 0 \leq t \leq T   \\
x_{0}&=\Phi =\{\Phi(\theta), ~\forall \theta \in [-\tau,0]\}
\end{aligned}\right.
\end{equation}
where $x_{t}=\{x(t+\theta):-\tau \leq \theta \leq 0\}$ is regarded as an element $C\left([-\tau, 0] ; \mathbb{R}^{5}\right)$-valued stochastic process, equipped with the norm $\|\Phi\|=\sup _{-\tau \leq \theta \leq 0}|\Phi(\theta)|$, then one can apply the existence and uniqueness  \cite[theorems 2.2 p 150]{ref29} to the delay equation \eqref{eq:4} when the function $f$ and $g$ satisfy the local Lipschitz and the linear growth condition. However in our case, the coefficients of the system \eqref{eq:3} do not satisfy the linear growth condition (because for example the incidence is nonlinear $S(t)(\beta_1E(t)+\beta_2I(t))$) but only the local Lipschitz, and so the solution of the system (\ref{eq:3}) exist but can explode at a finite time. Hence, we need to prove that the solution of the system (\ref{eq:3}) is positive and global and to do that we are inspired by the proof in   \cite[Theorem 2.1]{ref30}.
$f:  \left[0, T\right]\times C\left([-\tau, 0] ; \mathbb{R}^{5}\right) \rightarrow \mathbb{R}^{5} ~ \text { and } ~ g: \left[0, T\right]\times C\left([-\tau, 0] ; \mathbb{R}^{5}\right) \rightarrow \mathbb{R}^{ 5} \times \mathbb{R}^{ 5}
$

\begin{theorem}\label{thm : 3.1}
Under the condition $0 < \varphi \tau < e^{-1}$, the system (\ref{eq:3}) has a unique solution $(S(t),E(t),I(t),R(t),D(t))$ on $t\geq -\tau$, and the solution will remain in $\Gamma$ with probability one.
\end{theorem}
\begin{proof}
Equation (\ref{eq:3}) satisfy the local Lipschitz condition, so the system has a unique local solution on $t\in [-\tau, \tau_e)~a.s$ (\ref{THE6}), where $\tau_e$ is the explosion time. Our aim is to show that, this solution is global i.e $\tau_e=+\infty$. As the variable D does not appear in the first four equations of the system, it is sufficient to analyze the behavior of the solutions of the first four equations of the system. Let  $k_0>0$, be sufficiently large such that  $S(\theta),~E(\theta),~I(\theta)$ and $R(\theta)$, where $\theta \in [-\tau,0] $, are lying in
the interval $\left[\frac{1}{k_0},k_0\right]$. For each integer $k\geq k_0$, define the stopping time
$$
\tau_k = \inf \{t \in [0, \tau_e], min\{S(t),E(t),I(t),R(t)\}\leq \frac{1}{k}~or~max\{S(t),E(t),I(t),R(t)\}\geq k \},
$$
with the convention  $\inf\phi=\infty$ (where $\phi$ is the empty set). Since, $\tau_{k}$ is increasing as $k \rightarrow \infty$, we define $\tau_{\infty}=$ $\lim _{k \rightarrow \infty} \tau_{k}$, then we have $\tau_{\infty} \leq \tau_{e}$ a.s. Therefore, it is sufficient to show that $\tau_{\infty}=\infty$ a.s. to deduce that $\tau_{e}=\infty$ a.s. and $(S(t), E(t), I(t), R(t),D(t)) \in \Gamma$ a.s. for all $t \geq-\tau$. We proceed by the absurd and we assume that there is a pair of constants $\epsilon \in(0,1)$ and $\widetilde{T}>0$ such that 
$P\left\{\tau_{\infty} \leq \widetilde{T}\right\}>\epsilon$. 
Then, for all  $k \geq k_{0}$, we have
\begin{equation}\label{eq:Pr}
 P\left\{\tau_{k} \leq \widetilde{T}\right\} \geq \epsilon.   
\end{equation}
Now, we define a $\mathcal{C}^{2}$-function $\mathcal{V}: \mathbb{R}_+^4 \rightarrow \mathbb{R}_{+}$ as follows:
$$
\begin{aligned}
\mathcal{V}(S(t), E(t), I(t), R(t))=& S(t)-1-\ln (S(t))+E(t)-1-\ln(E(t)) +I(t)-1-\ln(I(t)) +  R(t)-1-\ln(R(t)).
\end{aligned}
$$
By applying the Ito's formula we get
\begin{equation}\label{eq:ly}
\begin{aligned}
d\left(\mathcal{V}(S(t), E(t), I(t), R(t))\right)=& \mathcal{L} \mathcal{V} d t-\eta_1\left(S(t)-1\right)E(t) d W(t)+\eta_1 (E(t)-1)S(t) d W_1(t)+ \eta_2d W_2(t)\\
&+ \eta_3d W_3(t)+ \eta_4d W_4(t),
\end{aligned}    
\end{equation}
where $\mathcal{L} \mathcal{V}: \mathbb{R}_{+}^{4} \rightarrow \mathbb{R}_{+} $ is the drift part defined by 
$$
\begin{aligned}
\mathcal{L} \mathcal{V}&=\left(1-\frac{1}{S(t)}\right)(\Lambda+\varphi R(t-\tau)e^{-\mu \tau}-\beta_1S(t)E(t)-\beta_2S(t)I(t)-\mu S(t))+\frac{1}{2}\eta_1^2 E^2(t) +\frac{1}{2}\eta_2^2\\
&+\left(1-\frac{1}{E(t)}\right)(\beta_1S(t)E(t)+\beta_2S(t)I(t)-(\mu +\delta)E(t)+\frac{1}{2} \eta_1^2 S^2(t)+\frac{1}{2}\eta_3^2 \\
&+\left(1-\frac{1}{I(t)}\right)\left(  \delta E(t)- \left(\sigma+\mu\right) I(t)\right)+\frac{1}{2}\eta_4^2 +\left(1-\frac{1}{R(t)}\right)\left( \left( 1-\alpha\right)\sigma I(t)- \varphi  R(t-\tau)e^{-\mu\tau}-\mu R(t)\right)\\
&=\Lambda -\mu (S(t)+E(t)+I(t)+R(t)) -\frac{\Lambda}{S(t)} -\frac{\varphi R(t-\tau)e^{-\mu \tau}}{S(t)} +\beta_1E(t)+\beta_2I(t)+4\mu\\
&+\frac{1}{2}\eta_1^2E^2(t)- \beta_1 S(t)-\beta_2 S(t)\frac{I(t)}{E(t)}+\delta +\frac{1}{2} \eta_1^2 S^2(t) -  \delta \frac{(E(t)}{I(t)} + \sigma\\
& -\alpha \sigma I(t) - \left( 1-\alpha\right)\sigma \frac{I(t)}{R(t)}+ \varphi \frac{R(t-\tau)}{R(t)} e^{-\mu\tau}+\frac{1}{2}(\eta_2^2+\eta_3^2+\eta_4^2) \\
&\leq \Lambda +\beta_1E(t)+\beta_2I(t)+4\mu +\frac{1}{2}\eta_1^2 E^2(t) +\delta +\frac{1}{2} \eta_1^2 S^2(t)  + \sigma+ \varphi \frac{R(t-\tau)}{R(t)} e^{-\mu\tau}+\frac{1}{2}(\eta_2^2+\eta_3^2+\eta_4^2) \\
\end{aligned}
$$
Or, we can see that, $$
d N(t)=\left(\Lambda-\mu N(t)-\alpha \sigma I(t)\right) d t
$$
Then, for $\left(S(t),E(t), I(t), R(t),D(t)\right) \in \Gamma$, the equation (\ref{S}) remains true on the event $\{-\tau \leq t \leq \tau_k\}$. Since, $S(t),E(t), I(t)$ and $ R(t)$ are non-negative, we have for all $-\tau \leq t \leq \tau_k$,
\begin{equation}\label{SS}
\frac{\Lambda}{\alpha \sigma+\mu} \leq N(t) \leq \frac{\Lambda}{\mu} \quad \text{and} \quad S(t),E(t), I(t), R(t) \leq \frac{\Lambda}{\mu}.
\end{equation}
Putting this result in the computation of $\mathcal{L} \mathcal{V}$ above, we obtain on the event $\{-\tau \leq t \leq \tau_k\}$
$$
\begin{aligned}
\mathcal{L} \mathcal{V} &\leq \Lambda +(\beta_1+\beta_2)\frac{\Lambda}{\mu}+4\mu+\eta_1^2(\frac{\Lambda}{\mu})^2 +\delta  +  \sigma +\frac{1}{2}(\eta_2^2+\eta_3^2+\eta_4^2)+\varphi \frac{R(t-\tau)}{R(t)} e^{-\mu\tau} \\
\end{aligned}
$$
Moreover, under the condition $0 < \varphi \tau < e^{-1}$,  as $\Phi_4 > 0 $, then there exist $C > 0$ such that $\Phi_4(t) \geq Ce^{\lambda t}$ on the event $\{-\tau \leq t < \tau_k\}$. According to the lemma (\ref{lemma0}) and comparison theorem, we get for all $t \in [0, \widetilde{T}) $, $R(t) \geq Ce^{\lambda \widetilde{T}} > 0$, then there exist positive constants $ C_1$ and $C_2$ such that for all $\{-\tau \leq t < \tau_k\wedge \widetilde{T}\}$, we have 
$$
\begin{aligned}
\mathcal{L} \mathcal{V} &\leq C_1+ C_2 e^{-\lambda \widetilde{T}}
\end{aligned}
$$
Now, as the diffusion coefficient of (\ref{eq:ly}) is bounded on $t\leq \tau_k \wedge \widetilde{T}$, integrating and taking the expectation on both sides yields 
\begin{equation} \label{eq:ex}
\begin{aligned}
&E \mathcal{V}\left(S\left(\tau_{k} \wedge \widetilde{T}\right), E\left(\tau_{k} \wedge \widetilde{T}\right), I\left(\tau_{k} \wedge \widetilde{T}\right), R\left(\tau_{k} \wedge \widetilde{T}\right), D\left(\tau_{k} \wedge \widetilde{T}\right)\right) \leq  \mathcal{V}(S(0), E(0), I(0), R(0))+ \left(C_1+ C_2 e^{-\lambda \widetilde{T}}\right) \widetilde{T} .
\end{aligned}
\end{equation}

Let $\Omega_{k}=\left\{\tau_{k} \leq \widetilde{T}\right\}$, for $k \geq k_{0}$ and in view of (\ref{eq:Pr}), we obtain $P\left(\Omega_{k}\right) \geq \epsilon$ such that, for every $\omega \in \Omega_{k}$, there is at least one component of $(S\left(\tau_{k}, \omega \right), E\left(\tau_{k}, \omega\right), I\left(\tau_{k}, \omega\right), I\left(\tau_{k}, \omega\right), R\left(\tau_{k}, \omega\right))$ equaling either $k$ or $\frac{1}{k}$, then 
$$ \mathcal{V}\left(S\left(\tau_{k} \wedge \widetilde{T}\right), E\left(\tau_{k} \wedge \widetilde{T}\right), I\left(\tau_{k} \wedge \widetilde{T}\right), R\left(\tau_{k} \wedge \widetilde{T}\right)\right) \geq (k-1-ln(k)) \wedge (\frac{1}{k}-1+ln\left(k\right) )$$
According to (\ref{eq:ex}), we get
$$
\begin{aligned}
 \mathcal{V}(S(0), E(0), I(0), R(0))+\left(C_1+ C_2 e^{-\lambda \widetilde{T}}\right) \widetilde{T} & \geq E\left[1_{\Omega_{k}} \mathcal{V}\left(S\left(\tau_{k}\right), E\left(\tau_{k}\right), I\left(\tau_{k}\right), R\left(\tau_{k}\right)\right)\right]\\& \geq \epsilon (k-1-ln(k)) \wedge (\frac{1}{k}-1+ln\left(k\right))\\
\end{aligned}
$$
where $1_{\Omega_{k}}$ represents the indicator function of $\Omega_{k}$. Letting $k \rightarrow \infty$ yields
$$
\infty>  \mathcal{V}(S(0), E(0), I(0), R(0))+\left(C_1+ C_2 e^{-\lambda \widetilde{T}}\right)\widetilde{T}=\infty,
$$
which leads to a contradiction. It can be concluded that $\tau_{\infty}=\infty$ a.s., which proves the theorem.

\end{proof}
Now, we introduce the following theorem that gives a sufficient condition for the extinction of the disease. 

\begin{theorem} \label{thm : 3.2}
Under the condition $0 < \varphi \tau < e^{-1}$, let $(S(t), E(t), I(t), R(t),D(t)) \in \Gamma$ be the solution of system (\ref{eq:3}).
If 
\begin{enumerate}
    \item $\frac{\beta_1^2}{2\eta_1^2}+\frac{\Lambda}{\mu}\beta_2-\frac{1}{2}\left(\eta_3^2+\eta_4^2\right)-2\mu -\sigma <0 $, \label{COND1}\\
    \item $\left[\frac{\Lambda}{2\mu}(\beta_2+\beta_1)-\mu - \frac{\sigma}{2}\right]^2 < \left(\frac{\beta_1^2}{2\eta_1^2} -\frac{1}{2}\eta_3^2 -\mu \right)\left(\beta_2\frac{\Lambda}{\mu}-(\sigma+\mu)-\frac{1}{2}\eta_4^2\right) $,\label{COND2}
\end{enumerate}
then $E(t)$ and $I(t)$ tends to zero exponentially with probability one, i.e.
\begin{equation*} 
\begin{split}
\limsup_{t \rightarrow +\infty}\frac{\ln{(E(t)+I(t))}}{t} &< 0
\end{split}.
\end{equation*}
\end{theorem}
\begin{proof}
By applying Ito’s formula, we get $d\ln\left( E(t)+I(t)\right)=$
\begin{align*}
&  \frac{1}{E(t)+I(t)}\left( S(t)\left(\beta_1E(t)+\beta_2I(t)\right))-\mu(E(t)+I(t))-\sigma I(t) \right)dt - \frac{1}{2} \left(\frac{\eta_1 S(t)E(t)}{E(t)+I(t)}  \right)^2dt\\
&- \frac{1}{2} \left(\frac{\eta_3 E(t)}{E(t)+I(t)}  \right)^2dt - \frac{1}{2} \left(\frac{\eta_4 I(t)}{E(t)+I(t)}  \right)^2dt+ \frac{\eta_1 S(t)E(t)}{E(t)+I(t)}dW_1(t) + \eta_3 \frac{E(t)}{E(t)+I(t)}dW_3(t)+ \eta_4 \frac{I(t)}{E(t)+I(t)}dW_4(t)\\
=&\left(\frac{1}{E(t)+I(t)}\right)^2\left[ (\beta_1+\beta_2) S(t)I(t) E(t)+\beta_2 S(t)I^2(t)+ \left(\beta_1S(t)- \frac{1}{2} \eta_1^2 S^2(t)\right)E^2(t)-\mu E^2(t)-\mu E(t)I(t)-\sigma I(t)E(t) \right.\\
&\left. -\mu E(t)I(t)-\mu I^2(t)-\sigma I^2(t)- \frac{1}{2} \eta_3^2 E^2(t) - \frac{1}{2} \eta_4^2 I^2(t) \right]dt + dM_1(t)
\end{align*}
with $M_1(t) = \displaystyle \int_0^t \frac{\eta_1 S(r)E(r)}{E(r)+I(r)}dW_1(r)+\displaystyle \int_0^t \eta_3 \frac{E(r)}{E(r)+I(r)}dW_3(r)+\displaystyle \int_0^t \eta_4 \frac{I(r)}{E(r)+I(r)}dW_4(r)$. On one hand, by (\ref{SS}), we have
$$ (\beta_1+\beta_2) S(t)I(t) E(t)+\beta_2 S(t)I^2(t)\leq (\beta_1+\beta_2) \frac{\Lambda}{\mu}I(t) E(t)+\beta_2 \frac{\Lambda}{\mu}I^2(t),$$
on the other hand, we have 
\begin{align*}
\left(\beta_1S(t)- \frac{1}{2} \eta_1^2 S^2(t)\right)E^2(t) &= \left(- \frac{1}{2} \eta_1^2\left(S(t)-\frac{\beta_1}{\eta_1^2}\right)^2 +\frac{\beta_1^2}{2 \eta_1^2} \right)E^2(t) \leq  \frac{\beta_1^2}{2 \eta_1^2}E^2(t).
\end{align*}
Using these results in $d\ln (E(t)+I(t)) $ we get
\begin{align*}
d\ln (E(t)+I(t)) &\leq \left(\frac{1}{E(t)+I(t)}\right)^2\left[  (\beta_1+\beta_2) \frac{\Lambda}{\mu}I(t) E(t)+\beta_2 \frac{\Lambda}{\mu}I^2(t) +\frac{\beta_1^2}{2 \eta_1^2}E^2(t) -\mu E^2(t)-\mu E(t)I(t)-\sigma I(t)E(t)\right.\\
&\left. -\mu E(t)I(t)-\mu I^2(t)-\sigma I^2(t) - \frac{1}{2} \eta_3^2 E^2(t)  - \frac{1}{2} \eta_4^2 I^2(t)\right]dt +  dM_1(t)\\
&= \left(\frac{1}{E(t)+I(t)}\right)^2 \left[ (E(t)~~I(t))\begin{pmatrix}
\frac{\beta_1^2}{2 \eta_1^2}-\frac{1}{2}\eta_3^2 -\mu &  \frac{\Lambda}{2\mu}(\beta_2+\beta_1)-\mu - \frac{\sigma}{2}\\[3mm]
\frac{\Lambda}{2\mu}(\beta_2+\beta_1)-\mu -\frac{\sigma}{2} &\beta_2\frac{\Lambda}{\mu}-(\sigma+\mu)-\frac{1}{2}\eta_4^2\\[3mm]
\end{pmatrix} \begin{pmatrix}
 E(t) \\[3mm]
 I(t)\\[3mm]
\end{pmatrix}\right]dt+  dM_1(t).
\end{align*}
Now, similarly, as in \cite{ref3b}, we consider the matrix
$
\begin{pmatrix}
\frac{\beta_1^2}{2 \eta_1^2}-\frac{1}{2}\eta_3^2 -\mu &  \frac{\Lambda}{2\mu}(\beta_2+\beta_1)-\mu - \frac{\sigma}{2}\\[3mm]
\frac{\Lambda}{2\mu}(\beta_2+\beta_1)-\mu -\frac{\sigma}{2} &\beta_2\frac{\Lambda}{\mu}-(\sigma+\mu)-\frac{1}{2}\eta_4^2\\[3mm]
\end{pmatrix}
$
which is negative-definite under the conditions the (\ref{COND1},\ref{COND2}). Therefore if we denote by $\lambda_{\max }$ the largest eigenvalue of the above matrix then
$$
\begin{aligned}
d\ln (E(t)+I(t))\leqslant & -\left|\lambda_{\max }\right| \frac{E^{2}(t)+I^{2}(t)}{\left(E(t)+I(t)\right)^{2}} d t +  dM_1(t).
\end{aligned}
$$
Using $ \left(x+y\right)^{2} \leqslant 2\left(x^{2}+y^{2}\right) $ 
and integrating the above equation from 0 to t we obtain
\begin{equation*} \label{EF1}
\begin{split}
\ln (E(t)+I(t)) &\leq  \ln (E(0)+I(0))-\frac{1}{2}\left|\lambda_{\max }\right| + M_1(t).
\end{split}
\end{equation*}
By, $\forall~t \geq -\tau ~S(t)\leq \frac{\Lambda}{\mu}$, it is easy to check that $\lim_{t\rightarrow +\infty}\frac{<M_1,M_1>_t}{t} < \infty~~a.s.$ Hence by the large number theorem for martingales (\ref{LEM5}), we obtain $
\lim_{t\rightarrow +\infty}\frac{M_1(t)}{t} =0~~a.s.$
Therefore, by dividing by $t$ and taking the limit superior, we obtain
\begin{equation*} \label{eq:11}
\begin{split}
\limsup_{t \rightarrow +\infty}\frac{\ln{(E(t)+I(t))}}{t} \leq &-\frac{1}{2}\left|\lambda_{\max }\right|<0.
\end{split}
\end{equation*}
This completes the proof.

\end{proof}
\begin{rem}\label{rem}
If $\eta_1^2 < \frac{\beta_1 \mu}{\Lambda}$, by writing $\left(\beta_1S(t)- \frac{1}{2} \eta_1^2 S^2(t)\right)E^2(t) =  \left(- \frac{1}{2} \eta_1^2\left(S(t)-\frac{\beta_1}{\eta_1^2}\right)^2 +\frac{\beta_1^2}{2 \eta_1^2} \right)E^2(t)$, we get
\begin{align*}
\left(\beta_1S(t)- \frac{1}{2} \eta_1^2 S^2(t)\right)E^2(t) &\leq  \left(- \frac{1}{2} \eta_1^2\left(\frac{\Lambda}{\mu}-\frac{\beta_1}{\eta_1^2}\right)^2 +\frac{\beta_1^2}{2 \eta_1^2} \right)E^2(t) = \left(\beta_1\frac{\Lambda}{\mu}- \frac{1}{2} \eta_1^2 \left(\frac{\Lambda}{\mu}\right)^2\right)E^2(t).
\end{align*}
Hence, the introduced matrix in the proof becomes
$\begin{pmatrix}
\beta_1\frac{\Lambda}{\mu}- \frac{1}{2} \eta_1^2 \left(\frac{\Lambda}{\mu}\right)^2-\frac{1}{2}\eta_3^2 -\mu &  \frac{\Lambda}{2\mu}(\beta_2+\beta_1)-\mu - \frac{\sigma}{2}\\[3mm]
\frac{\Lambda}{2\mu}(\beta_2+\beta_1)-\mu -\frac{\sigma}{2} &\beta_2\frac{\Lambda}{\mu}-(\sigma+\mu)-\frac{1}{2}\eta_4^2\\[3mm]
\end{pmatrix} $
and the result of the above theorem will be obtained under the new conditions
if \begin{itemize}
    \item [$1.bis$] $\frac{\Lambda}{\mu}\left(\beta_1+\beta_2\right)-\frac{1}{2}\eta_1^2\left(\frac{\Lambda}{\mu}\right)^2-\frac{1}{2}\left(\eta_3^2+\eta_4^2\right)-2\mu -\sigma <0, $ \label{COND11}\\
    \item [$2.bis$] $\left[\frac{\Lambda}{2\mu}(\beta_2+\beta_1)-\mu - \frac{\sigma}{2}\right]^2 < \left(\frac{\Lambda}{\mu} \beta_1 -\frac{1}{2}\eta_1^2\left(\frac{\Lambda}{\mu}\right)^2 -\frac{1}{2}\eta_3^2 -\mu \right)\left(\beta_2\frac{\Lambda}{\mu}-(\sigma+\mu)-\frac{1}{2}\eta_4^2\right). $\label{COND21}
\end{itemize}
\end{rem}
\begin{lemma} \label{lemma2}
Under the condition $0 < \varphi \tau < e^{-1}$, let $(S(t), E(t), I(t), R(t),D(t)) \in \Gamma$ be a solution of system (\ref{eq:3}). Then, we have 
\begin{equation} \label{INE2}
\int_0^tS(r)dr  \geq  \frac{\Lambda}{\mu}t- \frac{a_2}{\mu a_1}(\sigma + \mu)\int_0^t (E(r)+I(r)) dr - \frac{1}{a_1 \mu}  G(t) +\frac{1}{a_1 \mu} M_2(t)    
\end{equation}
where $M_2(t) = \int_0^t a_1\eta_2S(r)dW_2(r)+\int_0^ta_1\eta_3E(r)dW_3(r)+\int_0^ta_2\eta_4I(r)dW_4(r)$ and $G(t) = a_1(S(t)-S(0))+a_1(E(t)-E(0))+a_2(I(t)-I(0))+a_1\varphi e^{-\mu\tau} (\int_{t-\tau}^t R(r)dr-\int_{-\tau}^0 R(r)dr)$
with $a_1 = \delta + \sigma + \mu$ and $a_2 = \delta + \mu.$
\end{lemma}
\begin{proof}
 Let $(S(t), E(t), I(t), R(t),~D(t)) \in \Gamma$ be a solution of system (\ref{eq:3}), we have
\begin{align*}
\mathrm{d}\left(a_1S(t)+a_1E(t)+ a_2I(t)+a_1\varphi e^{-\mu\tau}\int_{t-\tau}^t R(r)dt\right) &=
 (a_1\Lambda - a_1\mu S(t)-a_2(\sigma +\mu)(E(t)+I(t)) + a_1\varphi e^{-\mu\tau} R(t))dt +dM_2(t)\\
 &\geq (a_1\Lambda - a_1\mu S(t)-a_2(\sigma +\mu)(E(t)+I(t)))dt +dM_2(t).
 \end{align*}
Integrating the above equation between 0 and t, we get
\begin{align*}
  G(t) &\geq a_1\Lambda t - a_1\mu \int_0^t S(r) dr  - a_2(\sigma +\mu)\int_0^t (E(r)+I(r)) dr  +M_2(t),
\end{align*}
which leads to the relation (\ref{INE2}).
\end{proof}
\begin{defn}
The system (\ref{eq:3}) is said to be persistent in the mean if $$ \displaystyle \liminf_{t \rightarrow \infty} \frac{1}{t} \int_{0}^{t}(E(r)+I(r)) d r>0~~ a.s.$$
\end{defn}
The following theorem gives some sufficient conditions ensuring the persistence of the disease. 
\begin{theorem} \label{thm : 3.6}
Under the condition $0 < \varphi \tau < e^{-1}$, let  $(S(t), E(t), I(t), R(t),D(t)) \in \Gamma$ be the solution of system (\ref{eq:3}).
Assume that  
\begin{equation}\label{COND3}
 \frac{\beta_1\beta_2}{\beta_1+\beta_2}  \frac{\Lambda }{\mu } - \frac{1}{2} \frac{\eta_1^2 \Lambda^2}{\mu^2} -\frac{1}{2} (\eta_3^2+ \eta_4^2) -(\sigma +\mu) \geq 0.   
\end{equation}
Then the disease will be persistent in the mean.
\end{theorem}
\begin{proof}

Similarly to theorem (\ref{thm : 3.2}), by applying Ito’s formula, we get 
\begin{align*}
 d\ln(E(t)+I(t)) &= \left[\frac{S(t) (\beta_1E(t)+\beta_2I(t))}{E(t)+I(t)} - \mu-\sigma \frac{I(t)}{E(t)+I(t)}  - \frac{1}{2} \eta_1^2 S^2(t)\left(\frac{E(t)}{E(t)+I(t)}\right)^2 - \frac{1}{2} \eta_3^2 \left(\frac{E(t)}{E(t)+I(t)}\right)^2\right.\\
 &\left.- \frac{1}{2} \eta_4^2 \left(\frac{I(t)}{E(t)+I(t)}\right)^2 \right]dt +dM_1(t)\\
\end{align*}
where $M_1(t) = \displaystyle \int_0^t \frac{\eta_1 S(r)E(r)}{E(r)+I(r)}dW_1(r)+\displaystyle \int_0^t \eta_3 \frac{E(r)}{E(r)+I(r)}dW_3(r)+\displaystyle \int_0^t \eta_4 \frac{I(r)}{E(r)+I(r)}dW_4(r)$.
In addition, using $ \displaystyle \frac{E(t)+I(t)}{\beta_1E(t)+\beta_2I(t)} \leq \frac{1}{\beta_1} + \frac{1}{\beta_2}$, $ \displaystyle \frac{E(t)}{E(t)+I(t)} \leq 1$ and $ \displaystyle \frac{I(t)}{E(t)+I(t)} \leq 1$  then we have
\begin{align*}
 d\ln(E(t)+I(t))&\geq \left[\frac{\beta_1\beta_2}{\beta_1+\beta_2}S(t) - (\sigma +\mu)  - \frac{1}{2} \eta_1^2 \left(\frac{\Lambda}{\mu}\right)^2 -\frac{1}{2} (\eta_3^2+ \eta_4^2)\right]dt + dM_1(t)  
\end{align*}
Integrating the above equation between 0 and t, we get
\begin{align*}
 \ln(E(t)+I(t)) &\geq \ln(E(0)+I(0)) +\left[\frac{\beta_1\beta_2}{\beta_1+\beta_2}\int_0^tS(r)dr- (\sigma +\mu)t   - \frac{1}{2} \eta_1^2 \left(\frac{\Lambda}{\mu}\right)^2t -\frac{1}{2} (\eta_3^2+ \eta_4^2)t\right] + M_1(t)
\end{align*}
Using Lemma (\ref{lemma2}), we obtain
\begin{align*}
 \ln(E(t)+I(t)) &\geq \left[\frac{\beta_1\beta_2}{\beta_1+\beta_2} \frac{\Lambda}{\mu}- (\sigma +\mu)   - \frac{1}{2} \eta_1^2 \left(\frac{\Lambda}{\mu}\right)^2 -\frac{1}{2} (\eta_3^2+ \eta_4^2) \right]t  - \frac{\beta_1\beta_2}{\beta_1+\beta_2} \frac{(\delta + \mu)(\sigma + \mu)}{\mu (\delta+\sigma +\mu)} \int_0^t E(r)+I(r)dr \\
 &+\ln(E(0)+I(0)) - \frac{\beta_1\beta_2}{\beta_1+\beta_2} \frac{1}{ \mu(\delta+\sigma+\mu)} G(t)+ M_1(t) + \frac{\beta_1\beta_2}{\beta_1+\beta_2}\frac{1}{ \mu(\delta+\sigma+\mu)} M_2(t) 
\end{align*}
Therefore, by  (\ref{SS}) and  the strong law of large numbers for martingales it is easy to check that $\displaystyle \lim_{t\rightarrow \infty} \frac{M_1(t)}{t}=0,~\lim_{t\rightarrow \infty} \frac{M_2(t)}{t}=0$ and $\displaystyle \lim_{t \rightarrow \infty} \frac{G(t)}{t} = 0$.
We complete the proof using Lemma \ref{LEM7} and condition (\ref{COND3}).
\end{proof}

\section{Numerical results}
\label{sec:4}

\begin{figure}[!h]
    \centering
    \includegraphics[scale=0.3]{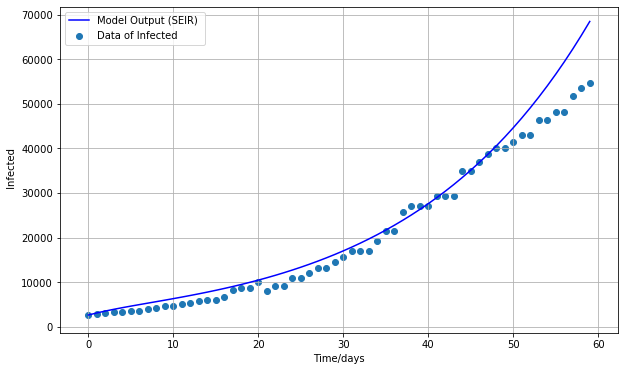}
    \includegraphics[scale=0.3]{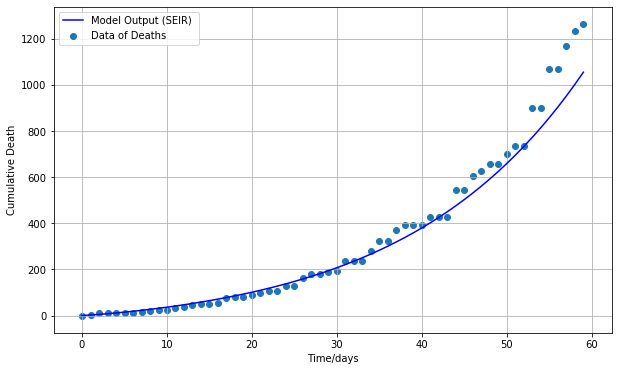}
    \caption{The result of the fitted value using the optimal parameters $\beta_1 =6.40541304000443e-09, ~\beta_2 =9.259947187897598e-09$, $\delta = 0.17571093$ and $\sigma =0.07885705$.}
    \label{F}
\end{figure}

The parameters and the initial condition used are shown in Table \ref{tab:my_label}. 
For model calibration, we utilize the system \ref{eq:0} and data from September $3–23, 2021$ (the start of the second wave) \footnote{https://covid19.who.int/WHO-COVID-19-global-data.csv}.
We assume that $\varphi = 0 $ and $\alpha = 0.01 $ during this time.
The natural death rate in Tunisia is $\Lambda = 205.52$ and the natural death rate  $\mu = 1.75 10^{-5}$. 
In order to estimate the transmission rates
 $\beta_1$ and $\beta_2$, recovered rate $\sigma$, and infected rate $\delta$, the mean square error between observed values and model simulations was minimized.
A genetic algorithm was used to determine the optimal
 The optimum was calculated using a genetic algorithm \footnote{https://github.com/rmsolgi/geneticalgorithm}. The model outputs of Death and Infected are contrasted with their actual data in Figure \ref{F}.

\begin{table}[!h]
\caption{Table of parameters}
    \label{tab:my_label}
    \centering
    \footnotesize  \begin{tabular}{cccc}
        \hline
        Parameter & Definition & Value & Reference \\
        \hline
        $N$   & The total size of the population & $11172177 $ &  \cite{ref17}\\
        $S(0)$   & The susceptible at time $0$ & $N(0)- E(0)- I(0) - R(0) - D(0) $ &  \cite{ref33}\\
        $E(0)$   & The exposed at time $0$ & $4000 $ &  \cite{ref33}\\
        $I(0)$   & The infected at time $0$ & $2629 $ &  \cite{ref33}\\
        $R(0)$   & The recovered at time $0$ & $0 $ &  \cite{ref33}\\
        $D(0)$   & The death at time $0$ & $0 $ &  \cite{ref33}\\
        $\Lambda$ & The newborn per unit of time & $205.51911751$& \cite{ref17}\\
        $\beta_1$ & The disease transmission coefficient of $E$ & $ 6.40541304000443e-09$& Fitted\\
        $\beta_2$ & The disease transmission coefficient of $I$ & $ 9.259947187897598e-09$ & Fitted\\
        $\mu$ & The natural death rate & $0.000017534$&  \cite{ref17} \\
        $\eta_1$ & The intensity of the perturbation of $\beta_1$ & $\cong \beta_0$ & Estimated\\
        $\eta_2$ &  Stochastic perturbations proportional to $S$ &$\in (0,1)$ & Estimated\\
        $\eta_3$ &  Stochastic perturbations proportional to $E$ &$\in (0,1)$ & Estimated\\
        $\eta_4$ &  Stochastic perturbations proportional to $I$ &$\in (0,1)$ & Estimated\\
        $\alpha$ & The death rate & $0.01$& \cite{ref33} \\
        $\delta$ & The rate at which exposed individuals become infectious & $ 0.17571093$& Fitted \\
        $\sigma$ & The recovered rate & $0.07885705$& Fitted \\
        $\tau$ & The period of temporary immunity & $60$, $96$, $201$, $360$ Days & Assumed \\
        $\varphi$ & The rate which individual loses his immunity& $\frac{1}{\tau e} - 0.00001 $  & Lemma \ref{lemma0}
    \end{tabular}
\end{table}

\begin{figure}[!h]
    \centering
    \includegraphics[scale=0.5]{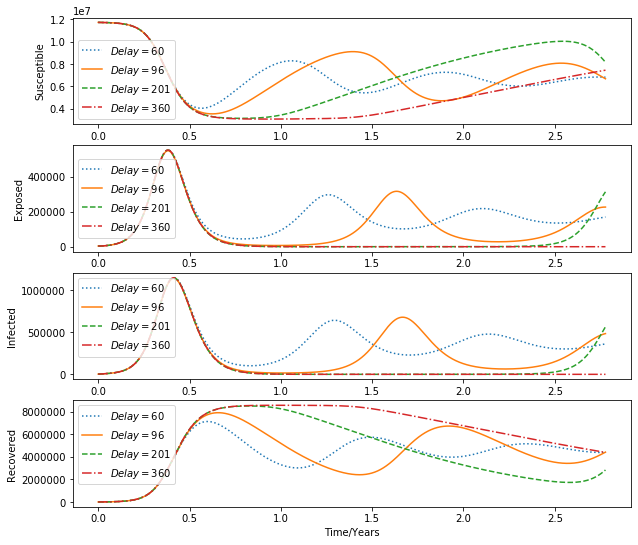}  
    \caption{SEIRDS simulation with varies delay, for roughly 3 years, for $60$, $96$, $201$ and $360$ days delay.
   }
\label{fig:Determinist Simulation}
\end{figure}


 We simulated  model \eqref{eq:1} for various values of immunity loss, $\tau= 60,~96,~201~\text{and}~360$ days and for $\varphi =  \frac{1}{\tau e}- 0.00001$ to satisfy the condition $0<\varphi \tau < e^{-1}$.
 Figure  \ref{fig:Determinist Simulation} shows a periodic epidemic due to the delay, $\tau$. We notice the amplitude of the oscillations and the period between waves increases with $\tau$. Indeed, if the loss of immunity occurs before the primary peak of the infection ($\tau \leq 96 $ days), this leads to more oscillations of the infection. On the opposite hand, if the loss of immunity occurs after the height of infection ($\tau \geq 201 $ days), subsequent waves are going to be distant (see Figures \ref{fig:Determinist Simulation}). Moreover, for $t$ large enough, the simulation will stabilize and converge to an endemic equilibrium.

\begin{figure}[!h]
    \centering
    \includegraphics[scale=0.5]{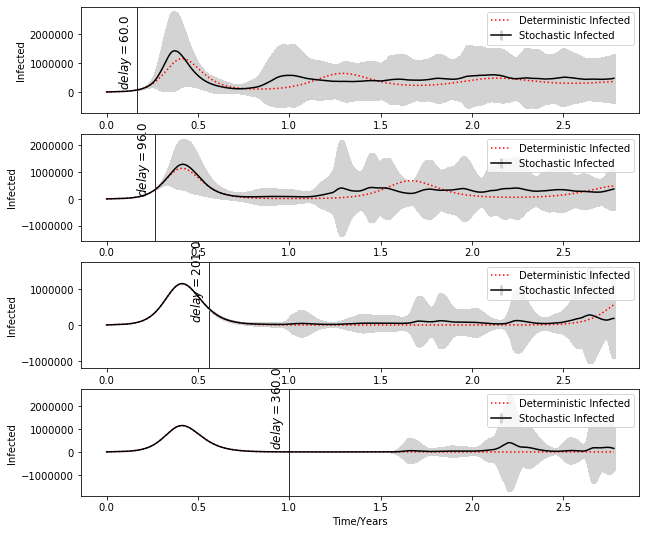}
    \caption{Infected simulation with varies delay, $60$, $96$, $201$, $360$ days for roughly 3 years.}
    \label{fig:Stochastic Vs determinist}
\end{figure}

We used the Euler-Maruyama method to discretize our stochastic model \eqref{eq:3}, and the solution of the deterministic model (\eqref{eq:0} with $\varphi=0$ as history for $t \leq \tau$ was used to simulate it:

\begin{equation}
\left\lbrace
\begin{aligned} 
S(t+\bigtriangleup t)&= S(t)+\left( \Lambda+ \varphi R(t-\tau)e^{-\mu\tau}- \beta_1 S(t) E(t)-\beta_2 S(t)I(t)-\mu S(t)\right) \bigtriangleup t
-\eta_1 S(t)E(t)\sqrt{\bigtriangleup t}\Psi_1 \\ &+\eta_2 S(t)\sqrt{\bigtriangleup t} \Psi_2\\
E(t+\bigtriangleup t) &= E(t) + \left( \beta_1 S(t) E(t)+\beta_2 S(t)I(t)-\left(\delta +\mu \right) E(t)\right) \bigtriangleup t + \eta_1 S(t) E(t) \sqrt{\bigtriangleup t}\Psi_1+\eta_3 E(t) \sqrt{\bigtriangleup t}\Psi_1\\
I(t+\bigtriangleup t) &= I(t)+ \left( \delta E(t)- \left(\sigma+\mu\right) I(t)\right) \bigtriangleup t +\eta_4 I(t)\sqrt{\bigtriangleup t}\Psi_1\\
R(t+\bigtriangleup t) &= R(t) + \left( \left( 1-\alpha\right)\sigma I(t)- \varphi  R(t-\tau) e^{-\mu\tau}-\mu R(t)\right) \bigtriangleup t\\
D(t+\bigtriangleup t) &= D(t) + \alpha \sigma I(t) \bigtriangleup t 
\end{aligned}
\right.
\label{eq:5}
\end{equation}
with 
$\Psi_1,~\Psi_2,\Psi_3,~\Psi_4~\text{are independent with low} ~\mathcal{N}(0,1),  \bigtriangleup t =  0.06  \text{ ~and~} \tau  \text{~is multiple of}  \bigtriangleup t.$ Afterward, for different values of $\tau$, we simulate $1000$ trajectories and plot the mean values of $I(t)$
and the confident interval ( see Figure \ref{fig:Stochastic Vs determinist}).
We observed repeated waves of infection that are smaller in size and occur less frequently, as well as the average settling towards the endemic equilibrium predicted by the deterministic model.
These effects are also accentuated in accordance with the delay. 

\begin{figure}[!h]
\centering
\includegraphics[scale=0.3]{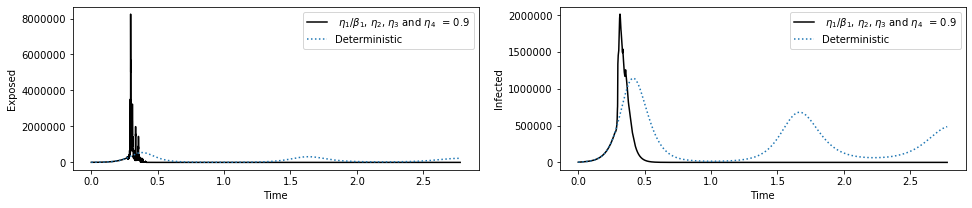}  
\caption{Simulation for Exposed $E$ and Infected $I$ with intensity $\eta_1/\beta_1=\eta_2=\eta_3=\eta_4= 0.9$.}
\label{fig:7e}
\end{figure}

\begin{figure}[!h]
    \centering
    \includegraphics[scale=0.3]{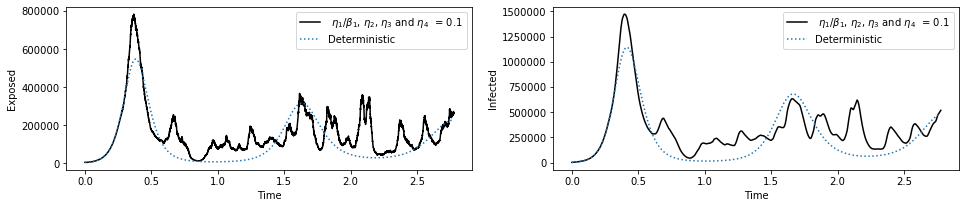}  
    \caption{Simulation for Exposed $E$ and Infected $I$ with intensity $\eta_1/\beta_1=\eta_2=\eta_3=\eta_4= 0.1$.}
    \label{fig:7a}
    \end{figure}

In Figures (\ref{fig:7e}) and (\ref{fig:7a})  we represent the simulation of exposed $E$ and infected $I$ versus $t$ for various values of $\eta_1,~\eta_2,~\eta_3$ and $~\eta_4$. According to Theorems (\ref{thm : 3.2}) and (\ref{thm : 3.6}) we can observe extinction or persistence of epidemic.

In example 1 we take $\beta_1=6.40\ 10^{-9}$, $\beta_2=9.25\ 10^{-9}$, $\delta=0.17$, $\sigma=0.07$ and we simulate the model (\ref{eq:5}) adding white noises values $\eta_{1}, \eta_{2}, \eta_{3}$, and $\eta_{4}$ such that the conditions (\ref{COND1}) and (\ref{COND2}) of Theorem (\ref{thm : 3.2}) hold. We observe the extinction of the disease in this situation as $E(t)$ and $I(t)$ tend to zero exponentially with probability one (see Figure  \ref{fig:7e}). However, in accordance with the deterministic model (\ref{eq:0}), we have $\mathcal{R}_{0}=4.47>1$, which indicates the existence of the endemic equilibrium. 

In example 2 we consider the same parameters of the example 1, $\beta_1$, $\beta_2$, $\delta$ and $\sigma$ and we simulate the model (\ref{eq:5}) choosing different values of white noises $\eta_{1}, \eta_{2}, \eta_{3}$ and $\eta_{4}$ that satisfy  assumptions of Theorem (\ref{thm : 3.6}). In this scenario, it is clear that the disease persists, as predicted by the deterministic model (see Figure \ref{fig:7a}). 

\section{Conclusion}
\label{sec:5}

This paper's main goal is to investigate how the loss of immunity affects the timing of the emergence of fresh waves.
We suggest that two SEIR/DS models be used to examine COVID-19 transmission.
A deterministic epidemic model is first taken into account. A mathematical model with five compartments and immunity loss was employed. The stochastic extended model SEIR/DS with noises is then taken into consideration.
Some model parameters are obtained from the literature, while the other parameters are determined by a genetic algorithm using actual daily data on COVID-19 cases in Tunisia who have died.

We prove that both models are appropriately formulated and make biological sense (see Theorem \ref{thm : 2.4} and \ref{thm : 3.1}).
Additionally, we investigate the stochastic model's behavior and provide conditions under which we have persistent or extinction of the epidemic (see Theorem \ref{thm : 3.2} and \ref{thm : 3.6}).
According to the numerical simulations, oscillations become more significant when immunity is lost before the first wave of infection.
And the subsequent waves will be further apart the more immunity is lost.



\section*{Acknowledgement}
This work was supported in part by the French Ministry for Europe and Foreign Affairs via the project “REPAIR COVID-19-Africa” coordinated by the Pasteur International  Network association  and by European Union's Horizon 2020 research and innovation program under grant agreement No. 883441 (STAMINA).
\section*{Authors’ contributions}
All authors read and approved the final manuscript.

\appendix
\section{}
\label{A}

\begin{lemma} \label{LEM2}
(\cite{ref15}, lemma 16.4: Comparison theorem page 215). Assume that $J$ and $D$ are open intervals in $\mathbb{R}$ and let $g \in C^{0,1-}(J \times D, \mathbb{R})$. Moreover, let $u \in C^{1}(J, D)$ be a solution of the differential equation $\dot{x}=g(t, x)$. Suppose $v \in C(J, D)$ and $\alpha \in J$ are such that
$$
v(\alpha) \leq u(\alpha) \quad \text { and } \quad D_{+} v(t) \leq g(t, v(t)), \quad \forall t \in J \cap[\alpha, \infty) .
$$
Then $v \leq u$ on $J \cap[\alpha, \infty)$.

Here, $
C^{0,1-}(J \times D, \mathbb{R}):=\{g: J \times D \rightarrow \mathbb{R} \mid g \in C(J \times D, \mathbb{R}), ~\text{and}~g~\text{is Lipschitz continuous with respect to}~x \in D\}
$
\end{lemma}
\begin{theorem} \label{THE3} 
(\cite{ref13}, Theorem $2.3$: Existence and Uniqueness on page 42). Let
\begin{equation}\label{xta}
    \left\lbrace
    \begin{aligned}
    &\dot{X}(t)=f\left(t,x_{t}\right)  \\
    &x_{0}=\Phi
    \end{aligned},
    \right.    
\end{equation}
Suppose $\Omega$ is an open set in $\mathbb{R} \times C, f: \Omega \rightarrow \mathbb{R}^{n}$ is continuous, and $f(t, \Phi)$ is Lipschitzian in $\Phi$ in each compact set in $\Omega$. If $\left( 0,\Phi\right) \in \Omega$, then there is a unique solution of Equation (\ref{xta}) through $\left( 0,\Phi\right)$.
\end{theorem}
\begin{lemma} \label{LEM5}
(\cite{ref29}, Theorem 3.4: Strong law of large numbers on page 12). Let $M=\left\{M_{t}\right\}_{t \geq 0}$ be a real-value continuous local martingale vanishing at $t=0$. Then
$$
\lim _{t \rightarrow \infty}\langle M, M\rangle_{t}=\infty \quad \text { a.s. } \Longrightarrow \lim _{t \rightarrow \infty} \frac{M_{t}}{\langle M, M\rangle_{t}}=0 \text {. a.s. }
$$
and also
$$
\limsup _{t \rightarrow \infty} \frac{\langle M, M\rangle_{t}}{t}<\infty \quad \text { a.s. } \Longrightarrow \lim _{t \rightarrow \infty} \frac{M_{t}}{t}=0 \text {. a.s. }
$$
\end{lemma} 
\begin{theorem} \label{THE6}
(\cite{ref29}, Theorem $2.8$ page 154) Assume that for every integer $n \geq 1$, there exists a positive constant $K_{n}$ such that, for all $t \geq 0$ and those $\varphi, \phi \in C\left(\{-\tau, 0] ; R^{d}\right)$ with $\|\varphi\| \vee$ $\|\phi\| \leq n$,
$$
|f(t,\varphi)-f(t,\phi)|^{2} \vee |g(t,\varphi)-g(t,\phi)|^{2} \leq K_{n}\|\varphi-\phi\|^{2}
$$
Then there exists a unique maximal local solution $x(t)$ to equation (\ref{eq:41}).
\end{theorem}

\begin{lemma} \label{LEM7} (Lemma 5.1. \cite{ref18b}).
Let $f \in C([0, \infty) \times \Omega,(0, \infty))$ and $G \in C([0, \infty) \times \Omega, \mathbb{R})$ such that $\lim _{t \rightarrow \infty} \frac{G(t)}{t}=0$ a.s. If for all $t \geq 0$
$$
\ln f(t) \geq \lambda_{0} t-\lambda \int_{0}^{t} f(s) d s+G(t) \text { a.s. }
$$
Then
$$
\liminf _{t \rightarrow \infty}\langle f(t)\rangle \geq \frac{\lambda_{0}}{\lambda} \text { a.s, }
$$
where $\lambda_{0} \geq 0$ and $\lambda>0$ are two real numbers.
\end{lemma}


\begin{thebibliography}{0}
\bibitem{ref1} Beretta, E., Kolmanovskii, V., \& Shaikhet, L. (1998). \emph{Stability of epidemic model with time delays influenced by stochastic perturbations}. Mathematics and Computers in Simulation, 45(3-4), 269-277.
\bibitem{ref2} Cao, W., Liu, M., \& Fan, Z. \emph{MS-stability of the Euler–Maruyama method for stochastic differential delay equations.} Applied Mathematics and Computation 159.1 : 127-135. (2004)
\bibitem{ref2a} Caraballo, T., El Fatini, M., El Khalifi, M., Gerlach, R., \& Pettersson, R. (2020). \emph{Analysis of a stochastic distributed delay epidemic model with relapse and gamma distribution kernel}. Chaos, Solitons \& Fractals, 133, 109643.

\bibitem{ref3}Chang, Z., Meng, X., \& Zhang, T. \emph{ A new way of investigating the asymptotic behaviour of a stochastic SIS system with multiplicative noise}. Applied Mathematics Letters, 87, 80-86.(2019).

\bibitem{ref3a} Cooke, K. L., \& Van Den Driessche, P. (1996). \emph{Analysis of an SEIRS epidemic model with two delays}. Journal of Mathematical Biology, 35(2), 240-260.

\bibitem{ref3b} Dalal, N., Greenhalgh, D., \& Mao, X. (2008). A stochastic model for internal HIV dynamics. Journal of Mathematical Analysis and Applications, 341(2), 1084-1101. 
 
\bibitem{ref4} Dan, J. M., Mateus, J., Kato, Y., Hastie, K. M., Yu, E. D., Faliti, C. E., ... \& Crotty, S. (2021). \emph{Immunological memory to SARS-CoV-2 assessed for up to 8 months after infection}. Science.

\bibitem{ref4aa} Driver, R. D., Sasser, D. W., \& Slater, M. L. (1973). \emph{The equation $x'(t)= ax (t)+ bx (t- \tau)$ with "small" delay.} The American Mathematical Monthly, 80(9), 990-995.

\bibitem{ref4a} El Fatini, M., Sekkak, I., Laaribi, A., Pettersson, R., \& Wang, K. (2020). \emph{A stochastic threshold of a delayed epidemic model incorporating Lévy processes with harmonic mean and vaccination.} International Journal of Biomathematics, 13(07), 2050069.
\bibitem{ref4b} El Fatini, M., Pettersson, R., Sekkak, I., \& Taki, R. (2020). \emph{A stochastic analysis for a triple delayed SIQR epidemic model with vaccination and elimination strategies.} Journal of Applied Mathematics and Computing, 64(1), 781-805.

\bibitem{ref5} Edridge, A. W., Kaczorowska, J., Hoste, A. C., Bakker, M., Klein, M., Loens, K., ..., and  van der Hoek, L. (2020). \emph{Seasonal coronavirus protective immunity is short-lasting}. Nature medicine, 26(11), 1691-1693.

\bibitem{ref6}Engbert, R., Rabe, M. M., Kliegl, R., \& Reich, S. (2021). \emph{Sequential data assimilation of the stochastic SEIR epidemic model for regional COVID-19 dynamics}. Bulletin of mathematical biology, 83(1), 1-16.

\bibitem{ref7}Fang, Y., Nie, Y., \& Penny, M.\emph{Transmission dynamics of the COVID‐19 outbreak and effectiveness of government interventions: A data‐driven analysis.} Journal of medical virology. (2020), 92(6), 645-659. 

\bibitem{ref8}Gao, N., Song, Y., Wang, X., \& Liu, J. \emph{Dynamics of a stochastic SIS epidemic model with nonlinear incidence rates}. Advances in Difference Equations, (1), 1-19. (2019).

\bibitem{ref9} Gray, A., Greenhalgh, D., Hu, L., Mao, X., \& Pan, J. (2011). \emph{A stochastic differential equation SIS epidemic model}. SIAM Journal on Applied Mathematics, 71(3), 876-902.

\bibitem{ref10}Gray, A., Greenhalgh, D., Mao, X., \& Pan, J.  \emph{The SIS epidemic model with Markovian switching.} Journal of Mathematical Analysis and Applications, 394(2), 496-516.(2012).

\bibitem{ref11}Greenhalgh, D., Moneim, I.A. \emph{SIRS epidemic model and simulations using different types of seasonal contact rate.} Syst. Anal. Model. Simul. 43(5), 573–600 (2003).

\bibitem{ref12}Gudbjartsson, D. F., Norddahl, G. L., Melsted, P., Gunnarsdottir, K., Holm, H., Eythorsson, E., ... \& Stefansson, K. (2020). \emph{Humoral immune response to SARS-CoV-2 in Iceland}. New England Journal of Medicine, 383(18), 1724-1734.

\bibitem{ref12a} Hadeler, K. P. (1993). \emph{Pair formation models with maturation period.} Journal of Mathematical Biology, 32(1), 1-15.

\bibitem{ref13}Hale, J. K. (2012). \emph{Theory of functional differential equations (Vol. 3)}. Springer Science \& Business Media.

\bibitem{ref14}He, S., Peng, Y., \& Sun, K. (2020). \emph{SEIR modeling of the COVID-19 and its dynamics.} Nonlinear dynamics, 101(3), 1667-1680.

\bibitem{ref14a} Hethcote, H. W., \& van den Driessche, P. (1995). \emph{An SIS epidemic model with variable population size and a delay.} Journal of mathematical biology, 34(2), 177-194.
\bibitem{ref14b} Hethcote, H. W., Lewis, M. A., \& Van Den Driessche, P. (1989). \emph{An epidemiological model with a delay and a nonlinear incidence rate.} Journal of mathematical biology, 27(1), 49-64.

\bibitem{ref15}Herbert Amann, \emph{Ordinary Differential Equations: An Introduction to Nonlinear Analysis (Degruyter Studies in Mathematics)}, Walter De Gruyter Inc, 1990.


\bibitem{ref16} Huang, A. T., Garcia-Carreras, B., Hitchings, M. D., Yang, B., Katzelnick, L. C., Rattigan, S. M., ... and Cummings, D. A. (2020). \emph{A systematic review of antibody mediated immunity to coronaviruses: kinetics, correlates of protection, and association with severity.} Nature communications, 11(1), 1-16.

\bibitem{ref17} Indicateurs du World-Factbook [archive] publié par la CIA.

\bibitem{ref18}Iwata, K., \& Miyakoshi, C. (2020). \emph{A simulation on potential secondary spread of novel coronavirus in an exported country using a stochastic epidemic SEIR model}. Journal of clinical medicine, 9(4), 944.
\bibitem{ref18b} Ji, C., Jiang, D., O’Regan, D.: \emph{Threshold behaviour of a stochastic SIR model}. Appl. Math. Comput.
38, 5067–5079 (2014)


\bibitem{ref19} Kermack, W. O., and McKendrick, A. G. \emph{A contribution to the mathematical theory of epidemics} (1927). Proceedings of the royal society of london. Series A, Containing papers of a mathematical and physical character, 115(772), 700-721.
\bibitem{ref19a} Kuang, Y. (Ed.). (1993). \emph{Delay differential equations: with applications in population dynamics}. Academic press.
\bibitem{ref21}L'Ecuyer, P.  \emph{Monte Carlo and Quasi-Monte Carlo Methods} Ed. Art B. Owen. Berlin: Springer, (2009).

\bibitem{ref22} Li, M. Y. (2018). \emph{An introduction to mathematical modeling of infectious diseases (Vol. 2).} Springer.

\bibitem{ref23}Liu, L., Meng, X.: \emph{Optimal harvesting control and dynamics of two-species stochastic model with delays}. Adv. Differ. Equ. 2017, 18 (2017)

\bibitem{ref24} Liu, M., Bai, C., \& Wang, K.  \emph{Asymptotic stability of a two-group stochastic SEIR model with infinite delays}. Communications in Nonlinear Science and Numerical Simulation, 19(10), 3444-3453.(2014).

\bibitem{ref26}Mahrouf, M., Boukhouima, A., Zine, H., Lotfi, E. M., Torres, D. F. M., \& Yousfi, N. (2021). \emph{Modeling and Forecasting of COVID-19 Spreading by Delayed Stochastic Differential Equations}. Axioms 2021, 10, 18.

\bibitem{ref27} Manou-abi, S., \& Balicchi, J. (2020). \emph{Analysis of the COVID-19 epidemic in french overseas department Mayotte based on a modified deterministic and stochastic SEIR model}. MedRxiv.

\bibitem{ref28}Mao, Xuerong. \emph{The truncated Euler–Maruyama method for stochastic differential equations.} Journal of Computational and Applied Mathematics 290 : 370-384. (2015)



\bibitem{ref29} Mao, X. (2007). \emph{Stochastic differential equations and applications}. Elsevier.

\bibitem{ref30} Mao, X., Marion, G., \& Renshaw, E. (2002). \emph{Environmental Brownian noise suppresses explosions in population dynamics.} Stochastic Processes and their Applications, 97(1), 95-110.

\bibitem{ref31} Miao, A., Zhang, J., Zhang, T., \& Pradeep, B. G.. \emph{Threshold dynamics of a stochastic model with vertical transmission and vaccination}. Computational and mathematical methods in medicine,  (2017).

\bibitem{ref32} Miao, A., Wang, X., Zhang, T., Wang, W., \& Pradeep, B. S. A.\emph{Dynamical analysis of a stochastic SIS epidemic model with nonlinear incidence rate and double epidemic hypothesis.}  Advances in Difference Equations, 2017(1), 1-27.(2017).


\bibitem{ref33} Novel Coronavirus (2019-nCoV) situation reports - World Health Organization (WHO)

\bibitem{ref34}Prado-Vivar, B., Becerra-Wong, M., Guadalupe, J. J., Márquez, S., Gutierrez, B., Rojas-Silva, P., ... \& Cárdenas, P. (2020). \emph{A case of SARS-CoV-2 reinfection in Ecuador}. The Lancet Infectious Diseases.

\bibitem{ref35}Rihan, F. A., Alsakaji, H. J., \& Rajivganthi, C. (2020). \emph{Stochastic SIRC epidemic model with time-delay for COVID-19.} Advances in difference equations, 2020(1), 1-20.

\bibitem{ref36}Shi, P., Cao, S., \& Feng, P. \emph{SEIR Transmission dynamics model of 2019 nCoV coronavirus with considering the weak infectious ability and changes in latency duration}. MedRxiv.(2020).

\bibitem{ref36a} Smith, H. L. (2011). An introduction to delay differential equations with applications to the life sciences (Vol. 57). New York: Springer.

\bibitem{ref37} Siggins, M. K., Thwaites, R. S., \& Openshaw, P. J. (2021). \emph{Durability of immunity to SARS-CoV-2 and other respiratory viruses}. Trends in Microbiology. 

\bibitem{ref38}Tillett, R. L., Sevinsky, J. R., Hartley, P. D., Kerwin, H., Crawford, N., Gorzalski, A., ... \& Pandori, M. (2021). \emph{Genomic evidence for reinfection with SARS-CoV-2: a case study.} The Lancet Infectious Diseases, 21(1), 52-58.

\bibitem{ref39} Van den Driessche, P., \& Watmough, J. (2002). \emph{Reproduction numbers and sub-threshold endemic equilibria for compartmental models of disease transmission}. Mathematical biosciences, 180(1-2), 29-48.

\bibitem{ref40}Wang, C., Horby, P. W., Hayden, F. G., \& Gao, G. F. (2020). \emph{A novel coronavirus outbreak of global health concern}. The lancet, 395(10223), 470-473.

\bibitem{ref41} Yang, X., Chen, L., \& Chen, J. (1996). \emph{Permanence and positive periodic solution for the single-species nonautonomous delay diffusive models.} Computers \& Mathematics with Applications, 32(4), 109-116.

\bibitem{ref42}Zhang, X., Jiang, D., Hayat, T., \& Ahmad, B. \emph{Dynamics of a stochastic SIS model with double epidemic diseases driven by Lévy jumps}. Physica A: Statistical Mechanics and its Applications, 471, 767-777. (2017).
\bibitem{ref43} Carleton, T., Cornetet, J., Huybers, P., Meng, K. C., \& Proctor, J. (2021). Global evidence for ultraviolet radiation decreasing COVID-19 growth rates. Proceedings of the National Academy of Sciences, 118(1), e2012370118.
\bibitem{ref44} Saha, J., Mondal, S., \& Chouhan, P. (2021). Spatial-temporal variations in community mobility during lockdown, unlock, and the second wave of COVID-19 in India: A data-based analysis using google's community mobility reports. Spatial and Spatio-temporal Epidemiology, 39, 100442.
\bibitem{ref45}Kermack, W. O., \& McKendrick, A. G. (1932). Contributions to the mathematical theory of epidemics. II.—The problem of endemicity. Proceedings of the Royal Society of London. Series A, containing papers of a mathematical and physical character, 138(834), 55-83.
\bibitem{ref46}Kermack, W. O., \& McKendrick, A. G. (1933). Contributions to the mathematical theory of epidemics. III.—Further studies of the problem of endemicity. Proceedings of the Royal Society of London. Series A, Containing Papers of a Mathematical and Physical Character, 141(843), 94-122.
\bibitem{ref47}Cohn AC, Mahon BE, Walensky RP. One Year of COVID-19 Vaccines: A Shot of Hope, a Dose of Reality. JAMA. 2022;327(2):119–120. doi:10.1001/jama.2021.23962
\bibitem{ref48}Rolland, Y., Cesari, M., Morley, J.E. et al. COVID19 Vaccination in Frail People. Lots of Hope and Some Questions. J Nutr Health Aging 25, 146–147 (2021). https://doi.org/10.1007/s12603-021-1591-9
\end{thebibliography}







\end{document}